\documentclass{article}

\usepackage{amssymb}
\usepackage{graphicx}
\begin{document}

\begin{center}

{\Large\bf INSTABILITY OF SCALAR\\[5PT]
PERTURBATIONS IN A PHANTOMIC\\[5PT]
COSMOLOGICAL SCENARIO\\[5PT]}
\medskip

J.C. Fabris\footnote{e-mail: fabris@pq.cnpq.br. On leave of absence of
Departamento de F\'{\i}sica, Universidade Federal do Esp\'{\i}rito Santo,
29060-900, Vit\'oria, Esp\'{\i}rito Santo, Brazil}\\
Institut d'Astrophysique de Paris\\
98bis, Boulevard Arago, 75014 Paris, France
\medskip
\\
\medskip
D.F. Jardim \footnote{e-mail: dfarago@bol.com.br} and S.V.B. Gon\c{c}alves\footnote{e-mail: sergio.vitorino@pq.cnpq.br}\\
Departamento de F\'{\i}sica, Universidade Federal do Esp\'{\i}rito Santo,
29060-900, Vit\'oria, Esp\'{\i}rito Santo, Brazil
\medskip

\end{center}

\begin{abstract}
Scalar perturbations can grow during a phantomic cosmological
phase as the big rip is approached, in spite of the high
accelerated expansion regime, if the equation of state is such
that $\frac{p}{\rho} = \alpha < - \frac{5}{3}$. It is shown that
such result is independent of the spatial curvature. The perturbed
equations are exactly solved for any value of the curvature
parameter $k$ and of the equation of state parameter $\alpha$.
Growing modes are found asymptotically under the condition $\alpha
< - \frac{5}{3}$. Since the Hubble radius decreases in a phantom
universe, such result indicates that a phantom scenario may not
survive longtime due to gravitational instability.
\end{abstract}

PACS: 98.80.-k, 95.36.+x
\vspace{0.5cm}

In the ordinary theory of cosmological perturbation
\cite{weinberg,padmanabhan,brand} there are two fundamental
regimes defined from the notion of {\it Jean's length},
$\lambda_J$. Perturbations whose scales are such that $\lambda >
\lambda_J$ suffer gravitational instability since the
gravitational attraction dominates over the pressure reaction to
contraction; on the other hand, perturbations whose scales satisfy
the condition $\lambda < \lambda_J$, tends to oscillates due to
the effectiveness of the pressure opposition to the gravitational
collapse. In the relativistic version of the theory of
cosmological perturbation we must add a new relevant scale, given
by the Hubble radius. The Hubble radius define, in some sense, the
effective causal region, and consequently the region where the
effects of the microphysics phenomenon (responsible for the
pressure) may play a relevant r\^ole in the process of
gravitational collapse. Perturbations whose scales are much large
than the Hubble radius tends to become frozen, while for those
smaller than the Hubble radius, the pressure tends to produce a
damping effect. Moreover, if the spatial section of the metric
representing the universe is not flat a new scale appears which is
connected to the curvature parameter.
\par
In cosmology, concomitant with the interplay between the
gravitational attraction and the pressure resistance, there is a
supplementary relevant effect due to the expansion of the
universe, which acts as a friction term in the fundamental
equations of the perturbed quantities: the expansion leads to a
damping in the evolution of perturbations. In general, for very
large perturbations, that are connected with the large structures
existing in the universe, there are two modes: a growing or a
constant mode and a decreasing mode. However, if the pressure is
negative, and the expansion becomes accelerated, gravity may
become repulsive and we generally find only decreasing modes: the
perturbations are always damped and structures can not be formed.
\par
There are quite strong evidences that the universe today is in an
accelerated expansion phase \cite{spergel}. If this is true, the
material content of the universe must be dominated by an exotic
fluid whose pressure is negative. There are many candidates for
this exotic fluid, like cosmological constant, quintessence,
K-essence, Chaplygin gas, etc., each of them having its advantages
and disadvantages from the theoretical point of view. We address
the reader to recent reviews on the many dark energy models
existing in the literature \cite{sahni,durrer}. Such negative
pressure fluid may generate repulsive effects driving the
accelerated expansion. There are some claims that the
observational data favors a phantom fluid, that is, a fluid whose
negative pressure violates the dominant energy condition $p + \rho
\geq 0$ \cite{caldwell}. Representing such fluid by a
self-interacting scalar field, a phantom fluid requires a "wrong"
sign in the kinetic term. This may imply instability at quantum
level. However, interesting propositions have been made in what
concerns such quite exotic fluid, mainly in the context of ghost
condensation existing in some string configurations \cite{piazza}.
\par
Classically, one of the most striking feature of phantom fluids is
the fact that its energy density grows with the expansion. This is
consequence of the violation of the strong energy condition, and
it may lead to a future singularity in a finite future time, the
so-called {\it big rip}. This is, of course, a very undesirable
feature. But it has been shown \cite{sergio1} that in a single
fluid approximation, for a spatially flat universe, the scalar
perturbations can grow when the scales of the perturbation are
greater than the Hubble radius. Since, the Hubble radius
decreases, for a phantom dominated universe, the isotropy and
homogenous condition would not be satisfied anymore as the big rip
is approached, leading perhaps to the avoidance of this future
singularity, leaving a very inhomogeneous universe. This situation
may occur if the pressure is negative enough in order to satisfy
the condition $\frac{p}{\rho} = \alpha < - \frac{5}{3}$.
\par
In the present work we extend those result showing that phenomenon
of enhancing of the inhomogeneities in large scales is independent
of the spatial curvature, and that the critical point $\alpha = -
\frac{5}{3}$ is present in any class of homogeneous and isotropic
universe. In order to do so, we will solve the perturbed equations
for scalar modes for any value of $k$ and $\alpha$. An asymptotic
analysis will reveal the existence of critical behavior associated
to $\alpha = - \frac{5}{3}$.
\par
For a universe dominated by a fluid whose equation of state is given by $p = \alpha\rho$, the relevant equation is
\begin{equation}
\frac{a'^2}{a^2} + k = \frac{8\pi G}{3}\rho\,a^2 \quad , \quad \rho = \rho_0\,a^{-3(1 + \alpha)} \quad .
\end{equation}
The primes indicate derivations with respect to the conformal time
$\eta$, and $k$ is the spatial curvature parameter, $k = \pm1,0$.
The solution for this equation may be written in a unified form as
\begin{equation}
a(\eta) = a_0\biggr[\frac{1}{\sqrt{k}}\sin\biggr(\frac{1 + 3\alpha}{2}\sqrt{k}\eta\biggl)\biggl]^\frac{2}{1 + 3\alpha} \quad .
\end{equation}
For $k = 1$, this solution represents a universe that begins and end at a singularity at $a = 0$ for $1 \leq \alpha < - \frac{1}{3}$, while
for $\alpha < - \frac{1}{3}$ it represents a bouncing universe. On the other hand, for $k = - 1, 0$, it represents an ever expanding universe
for any value of $\alpha$, with the following characteristics: for $1 \geq \alpha > - \frac{1}{3}$, the expansion implies $0 \leq \eta < \infty$
corresponding, in terms of the cosmic time $t$, to $0 \leq t < \infty$; for $- \frac{1}{3} > \alpha$, the expansion implies $- \infty < \eta \leq 0$,
corresponding to $0 \leq t < \infty$ if $- \frac{1}{3} > \alpha \geq - 1$ and $- \infty < t \leq 0$ if $- 1 > \alpha$. This last feature
leads to the notion of big rip.
\par
To study the evolution of scalar perturbations, we use the gauge invariant formalism. For a perfect fluid the evolution of the scalar perturbation
is given by a single equation for the gravitational potential $\Phi$ \cite{brand,mukhanov}:
\begin{equation}
\label{perturbed} \Phi'' + 3(1 + \alpha)H\Phi' +
\biggr\{\alpha\,n^2 + 2H' + (1 + 3\alpha)(H^2 - k)\biggl\}\Phi = 0
\quad ,
\end{equation}
where $H = \frac{a'}{a}$ and $n^2$ is the eigenvalue of the Laplacian operator $\nabla^2\Phi = - n^2 \Phi$.
The perturbed equation can be recast under the following form
\begin{equation}
\label{equation}
(1 - z)z\Phi'' + \frac{7 + 9\alpha}{2(1 + 3\alpha)}(1 - 2z)\Phi' + \tilde n^2\Phi = 0 \quad ,
\end{equation}
where
\begin{equation}
z = \frac{1 + \cos(\sqrt{k}\theta)}{2} \quad , \quad \tilde n^2 = \frac{4k}{(1 + 3\alpha)^2}\biggr[\alpha n^2 - 2(1 + 3\alpha)k\biggl] \quad , \quad
\theta = \frac{1 + 3\alpha}{2}\eta \quad .
\end{equation}
\par
The solution of (\ref{equation}) can be represented under the form of hypergeometric functions. It reads in general, for any value of $k$ and $\alpha$, as follows:
\begin{eqnarray}
\Phi_n(\eta) = c\, {}_2F_1\biggr[A_+,A_-;B;z\biggl] + \nonumber\\
\bar c\,z^{1 - B}{ }_2F_1\biggr[A_+ - B + 1,A_- - B + 1;2 - B;z\biggl] \quad ,
\end{eqnarray}
where
\begin{equation}
A_\pm = \frac{1}{2}\biggr\{6\frac{1 + \alpha}{1 + 3\alpha} \pm \sqrt{36\frac{(1 + \alpha)^2}{(1 + 3\alpha)^2} + 4\tilde n^2}\biggl\} \quad , \quad
B = \frac{7 + 9\alpha}{2(1 + 3\alpha)} \quad ,
\end{equation}
and $c,\bar c$ are constants.
\par
The asymptotic analysis for the flat case $k = 0$ was made in
reference \cite{sergio1}. In this case, the hypergeometric's
function reduces to Bessel's functions. It has been shown that one
of the two modes remains constant, in the long wavelength limit,
for any value of $\alpha$. In the same limit, however, the other
mode decreases when $\alpha > - \frac{5}{3}$, but grows with time
when $\alpha < - \frac{5}{3}$. For $\alpha = - \frac{5}{3}$, both
modes are constants. In the case $k \neq 0$, such analysis, in
terms of large or small wavelength limit, is more involved since,
contrarily to the flat case, we have a scale given by the Hubble
radius and another scale given by the curvature. It becomes
easier, in this sense, and for our purpose more relevant, to
consider the behavior in the extremes of the time interval. In
order to perform this analysis, we must take into account some
convenient transformation properties of the hypergeometric
functions. In special, the following transformations will be
useful (see reference \cite{grad}):
\begin{eqnarray}
{}_2F_1(A,B;C;z) = \frac{\Gamma(C)\Gamma(C-A-B)}{\Gamma(C-A)\Gamma(C-B)}{}_2F_1(A,B;A+B-C+1;1-z) &+&\nonumber\\
(1-z)^{C-A-B}\frac{\Gamma(C)\Gamma(A+B-C)}{\Gamma(A)\Gamma(B)}{}_2F_1(C-A,C-B;C-A-B+1;1-z)\quad &;&\\
{}_2F_1(A,B;C;z) = \frac{\Gamma(C)\Gamma(B-A)}{\Gamma(B)\Gamma(C-A)}(-1)^Az^{-A}{}_2F_1\biggr(A,A+1-C;A+1-B;\frac{1}{z}\biggl) &+&
\nonumber\\
\frac{\Gamma(C)\Gamma(A-B)}{\Gamma(A)\Gamma(C- B)}(-1)^Bz^{-B}{}_2F_1\biggr(B,B+1-C;B+1-A;\frac{1}{z}\biggl) \quad &.&
\end{eqnarray}
Hence, we have the following asymptotic behaviors:
\begin{eqnarray}
z \rightarrow 0 \quad \Rightarrow \quad {}_2F_1(A,B;C;z) &\sim&  z^{1 - C} \quad ,\\
z \rightarrow 1 \quad \Rightarrow \quad {}_2F_1(A,B;C;z) &\sim& \frac{\Gamma(C)\Gamma(C-A-B)}{\Gamma(C - A)\Gamma(C - B)}(1 - z)^{C-A-B}
\nonumber\\
&+& \frac{\Gamma(C)\Gamma(A+B-C)}{\Gamma(A)\Gamma(B)} \quad ,\\
z \rightarrow \infty \quad \Rightarrow \quad {}_2F_1(A,B;C;z) &\sim& \frac{\Gamma(C)\Gamma(B-A)}{\Gamma(B)\Gamma(C-B)}(-1)^A\,z^{-B}\nonumber\\
&+&\frac{\Gamma(C)\Gamma(A-B)}{\Gamma(A)\Gamma(C-B)}(-1)^B\,z^{-A} \quad .
\end{eqnarray}
\par
Using these expressions, we can determine the behavior of the
perturbations in the two different extremities of the time
interval, for each value of $k$.
\begin{itemize}
\item $k = 1$. In this case the conformal time interval is $0 \leq
\eta \leq \frac{2\pi}{1 + 3\alpha}$ for $\alpha > - \frac{1}{3}$
and $\frac{2\pi}{1 + 3\alpha} \leq \eta \leq 0$ for $\alpha < -
\frac{1}{3}$. Using the asymptotic expressions written above, we
find the following behaviors: for $\alpha > - \frac{1}{3}$, there
is initially two decreasing modes and, as the universe approaches
the big crunch at $\eta = \frac{2\pi}{1 + 3\alpha}$, there are a
constant mode and a growing mode; for $- \frac{1}{3} > \alpha > -
\frac{5}{3}$ there is initially, during the contraction phase, a
growing mode and a constant mode, and as the scale factor diverges
in the other asymptotic, there is a constant mode and a decreasing
mode; for $\alpha = - \frac{5}{3}$, both modes are constant at the
beginning of the contraction phase and at the end of the expansion
phase; for $\alpha < - \frac{5}{3}$, there is a constant mode and
a decreasing mode at the universe begins to contract, and there
are two increasing modes as the universe approaches the big rip.

\item $k = - 1$. The range of the conformal time is $0 \leq \eta <
\infty$ for $\alpha < - \frac{1}{3}$ and $- \infty < \eta \leq 0$
for $\alpha < - \frac{1}{3}$. The open case is more involved
because, in opposition to the closed universe, the asymptotic
behavior of the modes depends on the scale of the perturbation.
Let us consider the situation where the eigenvalues of the
Laplacian operator is null. Hence, when $\alpha > - \frac{1}{3}$
there is initially and in future infinity two decreasing modes.
For $- \frac{1}{3} < \alpha < - \frac{5}{3}$ there is initially
two decreasing modes, but in the future infinity there is a
constant mode besides a decreasing mode; the case $\alpha = -
\frac{5}{3}$ differs from the preceding one by the fact that in
future infinity both modes are constant. Finally, when $\alpha < -
\frac{5}{3}$, both modes are initially decreasing but they become
growing modes as the big rip is approached.
\end{itemize}
When $\alpha = - \frac{1}{3}$, in all cases, the same features observed in the flat universe are reproduced here, since for this particular equation of state the matter density scales as the curvature parameter in the Friedmann's equation.
\par
The main conclusion of the previous analysis is that a phantom
cosmological scenario is highly unstable against scalar
perturbations under the condition of an isotropic and homogeneous
background universe if the pressure is negative enough, that is
$\frac{p}{\rho} < - \frac{5}{3}$: the scalar perturbations grows
as the big rip is approached. As in the flat case, the Hubble
radius shrinks with time in the phantomic case. This means that
the large scale approximation becomes essentially valid
asymptotically for all perturbation scales in the phantomic case.
The analysis was made using a perfect fluid material content.
However, at large scales, it is expected that a more fundamental
representation, using for example, scalar fields, must give the
same results as the perfect fluid case \cite{sergio2}.
\par
A curious feature of the above results concerns the critical point
$\alpha = - \frac{5}{3}$. As far as we know, it does not
correspond to any energy condition (contrarily to $\alpha = -
\frac{1}{3}$ and $\alpha = - 1$). Also, it seems not related
neither to the Hubble parameter, nor to deceleration parameter, or
even to the statefinder parameters \cite{starobinsky}. However, it
seems to be a general critical point for the analysis of
perturbation, that is not reflected in the kinematical quantities
like those we have quoted above. A dynamical system analysis of
the background does not reveal any particular feature at $\alpha =
- \frac{5}{3}$ \cite{joel}: it is not a critical point for the
background. Moreover, in the perturbed equation (\ref{perturbed})
there is no explicit special structure for that particular value
of the parameter $\alpha$. The nature of this critical point must
still be cleared up. \vspace{0.5cm}
\newline
{\bf Acknowledgements:} We thank J\'er\^ome Martin for his
criticism and suggestions. We thank also CNPq (Brazil) and the
French-Brazilian scientific cooperation CAPES/COFECUB (project
number 506/05) for partial financial support.


\begin{thebibliography}{90}
\bibitem{weinberg} S. Weinberg, {\bf Gravitation and cosmology}, Wiley, New York (1972).
\bibitem{padmanabhan} T. Padmanabhan, {\bf Structure formation in the universe}, Cambridge University Press, Cambridge (1993).
\bibitem{brand} R. Brandenberger, H. Feldman and V. Mukhanov, Phys. Rep. {\bf 215}, 203 (1992).
\bibitem{spergel} D.N. Spergel et al, Astrophys. J. Suppl. {\bf 170}, 377 (2007).
\bibitem{sahni} V. Sahni, Lect. Notes Phys. {\bf 653}, 141 (2004).
\bibitem{durrer} R. D\"urrer and R. Maartens, {\it Dark energy and dark gravity}, arXiv:0711.0077.
\bibitem{caldwell} R. Caldwell, Phys. Lett. {\bf B545}, 23 (2002).
\bibitem{sergio1} J.C. Fabris and S.V.B. Gon\c{c}alves, Phys. Rev. {\bf D74}, 027301 (2006).
\bibitem{piazza} F. Piazza and S. Tsujikawa, JCAP {\bf 0407}, 004 (2007).
\bibitem{mukhanov} V. Mukhanov, {\bf Physical foundations of cosmology}, Cambridge University Press, Cambridge (2005).
\bibitem{grad} I.S. Gradshteyn and I.M. Rhyzik, {\bf Tables of integrals, series and products}, Academic Press, London (1994).
\bibitem{sergio2} J.C. Fabris, S.V.B. Gon\c{c}alves and N.A. Tomimura, Class. Quant. Grav. {\bf 17}, 2983 (2000).
\bibitem{starobinsky} V. Sahni, M. Sami, A.A. Starobinsky and U. Alam, JETP Lett. {\bf 77}, 201 (2003).
\bibitem{joel}  A.B. Batista, J.C. Fabris, S.V.B. Goncalves and J. Tossa, Int. J. Mod. Phys. {\bf A16}, 4527
(2001).

\end{thebibliography}
\end{document}